\documentclass[openacc]{rstransa}

\usepackage{graphicx}
\usepackage{epstopdf, epsfig}
\usepackage{amsmath}
\usepackage{color}
\usepackage{xcolor}

\newcommand{\bnabla}{\boldsymbol{\nabla}}
\newcommand{\la }{\left<}
\newcommand{\ra }{\right>}
\newcommand{\ez}{ {\bf e}_z}
\newcommand{\ex}{ {\bf e}_x}
\newcommand{\ey}{ {\bf e}_y}
\newcommand{\dz}{\mathrm{d}z}
\newcommand{\bu}{{\bf u}}
\newcommand{\RaP}{\mathrm{Ra_P}}
\newcommand{\ReRms}{\mathrm{Re}}
\newcommand{\Ra}{\mathrm{Ra}}
\newcommand{\Nu}{\mathrm{Nu}}
\newcommand{\prandtl}{\mathrm{Pr}}
\newcommand{\Ramax}{\mathrm{Ra_{max}}}
\newcommand{\Numax}{\mathrm{Nu_{max}}}
\newcommand{\cor}[1]{{#1}}

\begin{document}

\title{Velocity-informed upper bounds on the convective heat transport induced by internal heat sources and sinks}

\author{Vincent Bouillaut,
Beno\^it Flesselles,  
    Benjamin Miquel,
  S\'ebastien Auma\^itre,
 and Basile Gallet}

\address{$^{1}$ SPEC, CEA, CNRS, Universit\'e Paris Saclay, F-91191 Gif-sur-Yvette, France}

\corres{Basile Gallet\\
\email{basile.gallet@cea.fr}}

\begin{abstract}

Three-dimensional convection driven by internal heat sources and sinks (CISS) leads to experimental and numerical scaling-laws compatible with a mixing-length -- or `ultimate' -- scaling regime $\Nu \sim \sqrt{\Ra}$. However, asymptotic analytic solutions and idealized 2D simulations have shown that laminar flow solutions can transport heat even more efficiently, with $\Nu \sim \Ra$. The turbulent nature of the flow thus has a profound impact on its transport properties. In the present contribution we give this statement a precise mathematical sense. We show that the Nusselt number maximized over all solutions is bounded from above by $\text{const.}\times\Ra$, before restricting attention to 'fully turbulent branches of solutions', defined as families of solutions characterized by a finite nonzero limit of the dissipation coefficient at large driving amplitude. Maximization of $\Nu$ over such branches of solutions yields the better upper-bound $\Nu \lesssim \sqrt{\Ra}$. We then provide 3D numerical and experimental data of CISS compatible with a finite limiting value of the dissipation coefficient at large driving amplitude. It thus seems that CISS \cor{achieves the maximal  heat transport scaling} over fully turbulent solutions.
\end{abstract}


\maketitle

\section{Introduction}

Thermal convection refers to the fluid motion induced by the combined effect of vertical temperature gradients and gravity. The resulting motion enhances the heat transport from the warm to the cool fluid regions, as compared to the purely diffusive motionless state. A central question in both physics and applied mathematics of the Navier-Stokes equations is to determine this enhanced heat flux as a function of the strength of the driving
mechanism. The most common setup is the Rayleigh-B\'enard (RB) geometry, where a layer of fluid lies between a hot bottom plate and a cool top one, the two plates being maintained at constant temperature. In dimensionless form, the temperature difference between the two plates is quantified by the Rayleigh number $\Ra$. One then measures the heat flux across the cell, which after dividing by the purely diffusive heat flux of a motionless fluid layer yields the Nusselt number $\Nu$. The central question above then reduces to the determination of the asymptotic behavior of $\Nu$ as a function of $\Ra$, typically sought under the form $\Nu \sim \Ra^\gamma$ (leaving aside the Prandtl number $\prandtl$, defined as the ratio of the kinematic viscosity over the thermal diffusivity).

As initially proposed by Priestley and Malkus~\cite{priestley1954,malkus1954}, the emergent heat flux in the RB geometry is strongly restricted by the boundary layers adjacent to the top and bottom plates. This leads to the classical theory, characterized by a scaling exponent $\gamma=1/3$. This prediction departs from the mixing-length prediction of Spiegel~\cite{spiegel}: in the context of astrophysical fluids, where solid boundaries are irrelevant, Spiegel assumes that the heat flux is related to the temperature drop in a way that does not involve the tiny molecular diffusivities. This `diffusivity-free' argument leads to $\Nu \sim \sqrt{\Ra \, \Pr}$, that is, $\gamma=1/2$. While the RB studies do not agree on when and whether RB convection can achieve a heat transport exponent $\gamma$ greater than $1/3$ \cite{chavannePRL97,niemelaNAT00,chavannePoF01,Roche,hePRL12,Iyer2020,Doering2019,Doering2020}, they agree on the fact that the effective heat transport exponent is always significantly less than $1/2$, with experimental values in the range $\gamma \in [0.28 - 0.38]$. This makes it clear that the relationship between the heat flux and the temperature difference in experimental RB convection always involves the molecular diffusivities, either in a power-law or in a logarithmic fashion.

With the goal of circumventing the limitations of the RB setup, we recently introduced an alternate setup where convection is driven by internal heat sources and sinks (CISS). A combination of volumic sources and sinks deposits and removes heat directly inside the bulk turbulent flow, thus bypassing the throttling boundary layers of RB convection. In CISS, the heat flux is imposed and the vertical temperature drop is the emergent quantity that one measures or extracts from a numerical simulation. In dimensionless terms, the flux-based Rayleigh number $\RaP=\Nu \times \Ra$ is the control parameter, and the goal is to determine the Rayleigh number $\Ra$ for a given $\RaP$ (or, equivalently, to determine the Nusselt number $\Nu=\RaP/\Ra$). In the laboratory, the internal heat source corresponds to volumic absorption of light by a dyed fluid, while effective uniform cooling is realized by letting the body of fluid heat up -- the so-called 'secular heating' -- and focusing on the departure of the local temperature from the uniform drift (see \cite{lepotPNAS18,bouillautPNAS21} for details). We showed that this setup leads to the mixing-length regime of thermal convection, $\gamma=1/2$, provided the internal heating and cooling regions extend significantly beyond the thin boundary layers (\cite{bouillautJFM19}). The dependence in Prandtl number was further investigated through a suite of 3D direct numerical simulations (DNS), see \cite{miquelJFM20}. For a free-slip bottom boundary the Nusselt number satisfies Spiegel's mixing-length prediction in both $\Ra$ and $\Pr$, i.e., $\Nu \sim \sqrt{\Ra \, \Pr}$. The same holds for a no-slip bottom boundary condition at low $\Pr$, while persistent boundary layer corrections modify the behavior in $\Pr$ for large $\Pr$, with $\Nu \sim \Pr^{1/6}\, \sqrt{\Ra}$.

At the level of applied mathematics, one may hope to capture the mixing-length exponent $\gamma=1/2$ of CISS through the derivation of rigorous upper bounds on the Nusselt number, an approach pioneered by Howard, Busse and Doering \& Constantin~ \cite{howardJFM63,busseJFM69,doeringPRE96}.  For the RB setup, this approach leads to upper bounds of the form $\Nu \leq \text{const.}\times \Ra^{1/2}$, i.e., they are characterized by a diffusivity-free scaling exponent $\gamma=1/2$ greater than the exponent inferred from experiments and DNS. 
The natural question then is whether the exponent $\gamma=1/2$ of CISS corresponds to a maximization of the Nusselt number subject to simple constraints. We recently answered this question in the negative, computing upper bounds and exhibiting laminar asymptotic flow solutions of CISS characterized by a heat transport efficiency $\Nu \sim \Ra$, exceeding the mixing-length scaling exponent. Although the two are perfectly compatible at the mathematical level, there is a discrepancy between the behavior of upper bounds, which are \cor{saturated} by unstable (in 3D) laminar solutions  with a heat transport exponent $\gamma=1$, and the seemingly turbulent flows achieved in the laboratory and in 3D DNS, characterized by a mixing-length heat transport exponent $\gamma=1/2$. In other words, it seems that the turbulent nature of the flow has a profound impact on its transport properties. In the following, we confirm this statement in a precise mathematical sense.

Anticipating the precise definitions in section \ref{sec:VelInf}, we denote the velocity field as $\bu(x,y,z,t)$, the kinematic viscosity as $\nu$, the domain height as $H$, and a spatio-temporal average over the entire fluid domain with angular brackets. The dissipation coefficient ${\cal C}$ of the flow is then defined as:
\begin{equation}
{\cal C} =\frac{H\nu \la |\bnabla \bu|^2 \ra}{\la \bu^2 \ra^{3/2}} \, . \label{defC}
\end{equation}
\cor{`Turbulent dissipation' or `anomalous dissipation' refers to the singular limit of (\ref{defC}) as viscosity is lowered for constant large-scale forcing: even though viscosity appears at the numerator, turbulent flows develop stronger and stronger velocity gradients as viscosity is lowered, in such a way that (\ref{defC}) reaches a finite (strictly) positive limit as the Reynolds number goes to infinity~\cite{frischBOOK}. Such anomalous dissipation is used as the definition of a fully turbulent flow in the present study. More precisely,}
we define a 'fully turbulent branch of solutions' as a continuous family of solutions indexed by $\RaP$ that admits a finite nonzero limit of the dissipation coefficient for asymptotically large $\RaP$: 
\begin{equation}
\text{Fully turbulent branch of solutions} \iff \lim_{\RaP\to \infty} {\cal C} = {\cal C}_\infty  > 0. \label{eq:FTassumption}
\end{equation}
While we cannot prove the very existence of such turbulent branches of solutions, one can assume that such a branch of solutions exists -- an assumption referred to as the 'fully turbulent' assumption in the following -- and address motivational questions at the crossroad of applied mathematics and physics: 
\begin{itemize}
\item At the mathematical level, can one use information (or assumptions) about the velocity field to improve the upper bound on convective heat transport? Does CISS maximize the heat transport subject to simple constraints, when supplemented with the `fully turbulent' assumption?
\item At the physical level, can we validate this fully turbulent assumption using experimental and numerical data? Does the velocity field obey the free-fall scaling-law put forward by Spiegel to derive the mixing-length heat transport scaling-law~\cite{spiegel,Spiegel71}?
\end{itemize}

In section \ref{sec:VelInf}, we derive rigorous upper bounds on the Nusselt number in terms of the Rayleigh number and the dissipation coefficient. Assuming that a fully turbulent branch of solutions exists according to the definition (\ref{eq:FTassumption}) above, we show that the Nusselt number cannot increase faster that the square-root of the Rayleigh number over this branch of solutions. In other words, we show that the upper bound $\Nu \lesssim \sqrt{\Pr \, \Ra}$ holds for CISS if one restricts attention to fully turbulent solutions. By contrast, laminar flows can realize $\Nu \sim \Ra$, as established in \cite{miquelPRF19}, and we derive an upper bound on the Nusselt number valid for all flow solutions that reproduces this `laminar' scaling behavior: $\Nu \lesssim \Ra$.
In section \ref{sec:DNSExp} we turn to DNS and experimental realizations of CISS to establish the fully turbulent nature of the flow. We report 
experimental data pointing to fully turbulent dissipation and clearly discarding laminar-like dissipation. DNS allow for a careful study of the behavior of ${\cal C}$ with $\RaP$. The data point to a nonzero limiting value of the dissipation coefficient for increasingly large $\RaP$, in line with (\ref{eq:FTassumption}). We conclude in section \ref{sec:Disc}, the different results and data points being summarized in the schematic Figure~\ref{fig:bilan}, before making contact with the existing literature on CISS.

\section{Velocity-informed upper bounds}
\label{sec:VelInf}
\subsection{Boussinesq system of equations}

We consider a fluid layer inside a domain $(x,y,z)\in[0,L_x]\times[0,L_y]\times[0,H]$. Within the Boussinesq approximation, the dimensional equations governing the evolution of the velocity field $\bu(x,y,z,t)$ and the temperature field $\theta(x,y,z,t)$ read:
\begin{subequations}
\label{eq:dimensional_governing}
\begin{align}
\partial_t \bu + (\bu \boldsymbol{\cdot} \bnabla) \bu & =  -\bnabla p + \nu \bnabla^2 \bu + \alpha g \theta {\bf e}_z \, , \label{eq:NS}\\
\partial_t \theta + \bu \boldsymbol{\cdot} \bnabla \theta & =  \kappa \bnabla^2 \theta + \frac{P}{\rho C \ell_0} \left[ e^{-z/\ell_0} - \frac{\ell_0}{H} \left(1-e^{-H/\ell_0} \right)  \right] \, ,
\label{eq:HE}
\end{align}
\end{subequations}
where $\nu$ denotes the kinematic viscosity, $\alpha$ the thermal expansion coefficient, $g$ is gravity, $\kappa$ is the thermal diffusivity, $\rho$ is the mean fluid density and $C$ is the specific heat capacity. The last term in equation (\ref{eq:HE}) represents the internal heat sources and sinks. The precise $z$-dependence of this term is motivated by the experimental realization of CISS, see section~\ref{sec:Exp}. The first term inside the square bracket corresponds to the volumic absorption of light by a dyed fluid, where $P$ denotes the heat flux (in W.m$^{-2}$) provided by the light source in the form of visible light. This term decreases exponentially with height over an absorption length $\ell_0$ as a result of Beer-Lambert's law. The second term inside the square brackets corresponds to the effective uniform heat sink associated with the secular heating of the body of fluid. \cor{This uniform cooling term balances the radiative heat source on space average, i.e., the integral of the bracketed term over the domain height vanishes.}
We nondimensionalize the variables using $H$, $H^2/\kappa$ and $\nu \kappa /(\alpha g H^3)$ as length, time and temperature scales:
\begin{equation}
{\bf x}= H \tilde{\bf x} \, , \quad t= \frac{H^2}{\kappa} \tilde{t} \, , \quad \theta= \frac{\nu \kappa}{\alpha g H^3} \tilde{\theta} \, , \quad \bu = \frac{\kappa}{H} \tilde{\bu} \, .
\end{equation}
Dropping the tildes to alleviate notations, the dimensionless Boussinesq equations read:
\begin{subequations}
\label{eq:dimless_governing}
\begin{align}
\partial_t \bu + (\bu \boldsymbol{\cdot} \bnabla) \bu & = -\bnabla p + \Pr \bnabla^2 \bu + \Pr \theta {\bf e}_z \, , \label{equ}\\
\partial_t \theta + \bu \boldsymbol{\cdot} \bnabla \theta & =  \bnabla^2 \theta + \RaP \, S(z) \, .\label{eqtheta}
\end{align}
\end{subequations}
The dimensionless control parameters appearing in this set of equations are the Prandtl number and the flux-based Rayleigh number:
\begin{equation}
\Pr=\frac{\nu}{\kappa} \, , \qquad \RaP=\frac{\alpha g H^4 P}{\rho C \nu \kappa^2} \, . \label{eq:control_parameters}
\end{equation}
The heat source/sink function $S(z)$ is:
\begin{equation}
S(z)=\frac{e^{-z/\ell}}{\ell}-1+e^{-1/\ell} \, , \label{adimS}
\end{equation}
where ${\ell}=\ell_0/H$ denotes the dimensionless absorption length. \cor{The integral of $S(z)$ from $z=0$ to $z=1$ vanishes, because the uniform heat sink removes precisely the amount of heat input by the radiative heat source (over space average and per unit time).}

The set of equations (\ref{equ}-\ref{eqtheta}) is supplemented with the incompressibility constraint $\bnabla \cdot \bu =0$ and impermeable insulating boundary conditions at $z=0$ and $z=1$: 
\begin{equation}
w |_{z=0;1}=0 \, , \qquad \partial_z \theta |_{z=0;1} = 0 \, .
\label{eq:impermeable_insulating}
\end{equation}
We consider a no-slip bottom boundary:
\begin{equation}
    u|_{z=0} =v|_{z=0} = 0\, ,
    \label{eq:noSlip_bottom}
\end{equation}
while the top boundary condition is either free-slip or no-slip:
\begin{subequations}
\begin{align}
     u|_{z=1} &= v|_{z=1} = 0\, \label{eq:noSlip_top} \\ 
    \mathrm{or} \quad \left. \partial_z u\right|_{z=1} & = \left. \partial_z v\right|_{z=1} = 0 \, . \label{eq:stressFree_top}
\end{align}
\end{subequations}
Finally, we consider periodic boundary conditions in the horizontal directions.

We want to characterize the internal temperature fluctuations that emerge in this system. \cor{Integrating equation (\ref{eqtheta}) over the entire fluid domain indicates that the space average of $\theta$ is independent of time, because the vertical average of $S(z)$ vanishes: if heat is input and removed at the same rate inside an insulated container, the space-averaged fluid temperature remains constant. Without loss of generality and to alleviate notations, we thus assume in the following that $\theta$ is mean-zero initially and therefore at any subsequent time. As the first moment of the temperature field vanishes, the simplest nonzero measure of the emergent temperature fluctuations is arguably the second moment of the temperature field. One can thus quantify the emergent temperature fluctuations through a Rayleigh number $\Ra$ based on the root-mean-square temperature. In terms of the dimensionless variables, this leads to the simple definition $\Ra=\sqrt{\la \theta^2 \ra}$, where the angular brackets $\la \cdot \ra$ denote a space and time average. To define a Nusselt number, one can estimate the typical diffusive flux associated with the emergent temperature scale $\sqrt{\la \theta^2 \ra}$, had it been imposed at the large scale $H$ (that is, the diffusive flux that would arise if an equivalent temperature drop were imposed at large scale to a solid with the same thermal properties than the fluid). This equivalent diffusive flux is estimated simply as $\sqrt{\la \theta^2 \ra}$ in dimensionless form. We finally build the Nusselt number $\Nu$ by dividing the total input heat flux -- $\RaP$ in dimensionless form -- by this equivalent diffusive flux $\sqrt{\la \theta^2 \ra}$, which yields:}
\begin{eqnarray}
\Nu=\RaP / \Ra \, , \qquad  \Ra=\sqrt{\la \theta^2 \ra} \, . \label{defNuRa}
\end{eqnarray}
\cor{These definitions are well-suited for analysis and will be used extensively throughout this study. The reader might object that the Nusselt number defined above is not necessarily equal to unity in the diffusive state, a complication that would only modify the prefactors but not the scaling exponents of the bounds derived below.}

\cor{An alternate definition for the Nusselt could be based on the partitioning of the input potential energy into a diffusive and a convective contribution, see equation (\ref{eqtemp1}) below. The complication here is that one could imagine a situation where the emergent temperature fluctuations are large, albeit with a negligible diffusive vertical flux. This situation arises precisely for the analytical solution computed in Ref.~\cite{miquelPRF19}: the dominant temperature field in the expansion has a vanishing horizontal average at any height, and thus a vanishing (horizontally averaged) diffusive flux in the vertical direction at any height. A compromise between these two possible definitions for the Nusselt number is to base the estimate of the emergent temperature drop on the horizontally averaged squared temperature (instead of the horizontally averaged temperature). We thus consider an alternate temperature-based Rayleigh number built with the maximum in the vertical direction of the time and horizontal average of the squared temperature:
\begin{eqnarray}
\Ramax=\text{max}_{z\in[0;1]} \left\{ \sqrt{\overline{\theta^2}} \right\}  \, ,\label{eq:defRamax}
\end{eqnarray}
where the overbar denotes an average with respect to $x$, $y$ and $t$. A Nusselt number is then defined as the ratio of the input flux over the diffusive flux associated with the temperature estimate (\ref{eq:defRamax}) established over the entire height of the cell, leading to:
\begin{eqnarray}
\Numax=\RaP / \Ramax \, . \label{eq:defNumax}
\end{eqnarray}
The definitions of $\Ramax$ and $\Numax$ allow us to make better contact with the literature, as experimentalists typically measure the vertical temperature drop across a convection cell. In both laboratory experiments and DNS of radiatively driven convection, the maximum temperature is achieved at the bottom of the fluid domain and fluctuates moderately in the horizontal directions. $\Ramax$ is then a good proxy for the Rayleigh number based on the temperature drop between the bottom boundary and the bulk of the fluid (the latter definition for the Rayleigh number being the one used in previous experimental and numerical studies of this system, see Refs.~\cite{lepotPNAS18,bouillautJFM19,miquelJFM20}). Another desirable feature of $\Ramax$ and $\Numax$ is that, up to factors of two, they reduce to the standard definitions of the Rayleigh and Nusselt numbers when applied to the canonical fixed-temperature Rayleigh-Bénard setup.}

\cor{In the following, we derive upper bounds on $\Nu$ in terms of $\Ra$, but we stress the fact that all these bounds carry over to $\Numax$ and $\Ramax$. Indeed, from the definition} $\la \theta^2 \ra = \int_{0}^1 \overline{\theta^2} \dz$, we obtain $\la \theta^2 \ra \leq \text{max}_{z\in[0;1]} \{ \overline{\theta^2} \}$, hence:
\begin{eqnarray}
\Ra \leq \Ramax \, , \qquad \Numax \leq \Nu \, . \label{ineqRamax}
\end{eqnarray}
The upper bounds derived in the following are typically of the form $\Nu \leq c\times\Ra^{\gamma}$, with $\gamma$ a positive exponent and $c$ a prefactor. From the inequalities (\ref{ineqRamax}) one immediately obtains that these bounds carry over in terms of $\Ramax$ and $\Numax$, i.e., $\Numax \leq c \times \Ramax^{\gamma}$, the latter form being better-suited for comparison with experimental measurements.

\subsection{Bounding the heat flux in terms of the root-mean-square velocity}

Multiplying the temperature equation (\ref{eqtheta}) by $\theta$ before averaging over space and time yields, after a few integration by parts using the boundary conditions:
\begin{eqnarray}
\la |\bnabla \theta|^2 \ra = \RaP \int_0^1 S(z) \overline{\theta} \, \dz \, . \label{bilantheta2}
\end{eqnarray}

Similarly, multiplying the temperature equation (\ref{eqtheta}) by $z$ before averaging over space and time yields, after a few integration by parts using the boundary conditions:
\begin{eqnarray}
- \RaP \int_0^1 z S(z) \dz = \la w \theta \ra - \int_0^1 \partial_z \overline{\theta} \, \dz   \, . \label{eqtemp1}
\end{eqnarray}
We bound the second term on the right-hand side using the Cauchy-Schwarz inequality, equation (\ref{bilantheta2}) and \cor{Jensen's} inequality:
\begin{eqnarray}
 \left| \int_0^1 \partial_z \overline{\theta} \, \dz  \right|  \leq  \sqrt{\int_0^1 (\partial_z \overline{\theta})^2 \, \dz} & \leq &  \sqrt{\int_0^1 \overline{(\partial_z  {\theta})^2} \, \dz} \\
& & \nonumber  \leq \sqrt{\la |\bnabla \theta ^2| \ra} =  \sqrt{\RaP \int_0^1 S(z) \overline{\theta} \, \dz }  . \quad
\end{eqnarray}
Inserting this inequality into (\ref{eqtemp1}) and applying the Cauchy-Schwarz inequality to both terms on the right-hand side, we obtain:
\begin{eqnarray}
- \RaP \int_0^1 z S(z) \dz & \leq &  \la w \theta \ra + \sqrt{\RaP \int_0^1 S(z) \overline{\theta} \, \dz }  \\
&  \leq & \sqrt{\la w^2 \ra} \Ra +  \sqrt{\RaP \Ra} \left(  \int_0^1 S(z)^2  \, \dz  \right)^{1/4}
\end{eqnarray}
Dividing by the positive quantity $- \Ra \int_0^1 z S(z) \dz$ and using the relation $\RaP= \Ra \, \Nu$ leads to:
\begin{eqnarray}
\Nu - \Nu^{1/2} \frac{c_2}{c_1} - \frac{\sqrt{\la w^2 \ra} }{c_1} \leq 0 \, , \label{quadsqrtNu}
\end{eqnarray}
where $c_1$ and $c_2$ are the following positive constants (for a given geometry of the setup):
\begin{align}
c_1 & =  - \la zS \ra = \frac{1}{2}-\ell +e^{-1/\ell} \left(\frac{1}{2} +\ell \right) \, ,  \label{eq:c1}\\
c_2 & =  \la S^2 \ra^{1/4} = \left[-1-e^{-2/\ell}+2e^{-1/\ell}+\frac{1}{2\ell}-\frac{e^{-2/\ell}}{2\ell} \right]^{1/4} \label{eq:c2} \, .
\end{align}
Seeking the roots of the quadratic function of $\Nu^{1/2}$ on left-hand side of (\ref{quadsqrtNu}), one obtains the equivalent inequality:
\begin{eqnarray}
\Nu \leq \frac{1}{4} \left[ \frac{c_2}{c_1} + \sqrt{ \left(\frac{c_2}{c_1}  \right)^2 + 4 \frac{\la w^2 \ra^{1/2}}{c_1}}  \right]^2 \, . \label{boundwrms}
\end{eqnarray}
The right-hand side of the inequality above is an upper bound on the Nusselt number in terms of the root-mean-square vertical velocity.

\subsection{Bounding the heat flux in terms of the dissipation coefficient and the Rayleigh number}

We would like to bound the root-mean-square vertical velocity to turn the upper bound (\ref{boundwrms}) into a bound in terms of the Rayleigh number and the dissipation coefficient ${\cal C}$. Including the latter into the upper bound allows to readily derive upper bounds that apply to turbulent families of solutions, that is, to an hypothetical family of solutions characterized by a non-zero limiting value of the dissipation coefficient ${\cal C}$ as $\RaP$ (or $\Ra$) goes to infinity, see (\ref{eq:FTassumption}). We first derive the kinetic energy power integral, obtained by taking the dot product of (\ref{equ}) with $\bu$ before averaging over space and time. After a few integrations by parts, one obtains:
\begin{equation}
\la |\bnabla \bu|^2 \ra = \la w \theta \ra \, . \label{KEPI}
\end{equation}
In terms of the dimensionless variables, the definition (\ref{defC}) of the dissipation coefficient ${\cal C}$ becomes:
\begin{equation}
{\cal C} = \Pr \frac{\la |\bnabla \bu|^2 \ra}{\la \bu^2 \ra^{3/2}} \label{def:C}
\end{equation}
Using this definition, equation (\ref{KEPI}) and the Cauchy-Schwarz inequality, one obtains successively:
\begin{eqnarray}
\la w^2 \ra^{3/2} \leq \la \bu^2 \ra^{3/2} = \frac{\Pr}{{\cal C}} \la |\bnabla \bu|^2 \ra = \frac{\Pr}{{\cal C}} \la w \theta \ra \leq \frac{\Pr \, \Ra}{{\cal C}} \la w^2 \ra^{1/2} \, . \label{tempw}
\end{eqnarray}
Dividing across by $\la w^2 \ra^{1/2}$ and taking the square-root provides an upper bound on the root-mean-square vertical velocity:
\begin{eqnarray}
\la w^2 \ra^{1/2} \leq \sqrt{\frac{\Pr \, \Ra}{{\cal C}}} \, .
\end{eqnarray}
Substituting this inequality into (\ref{boundwrms}) leads to the following upper-bound ${\cal B}_1$ on the Nusselt number:
\begin{eqnarray}
\Nu \leq {\cal B}_1 = \frac{1}{4} \left[ \frac{c_2}{c_1} + \sqrt{ \left(\frac{c_2}{c_1}  \right)^2 + \frac{4}{c_1} \sqrt{\frac{\Pr \, \Ra}{{\cal C}}} }  \right]^2 \, . \label{boundB}
\end{eqnarray}
We are interested in the behavior of the upper-bound ${\cal B}_1$ as the Rayleigh number goes to infinity. For a given geometry (constant coefficients $c_1$ and $c_2$), the asymptotic behavior of the upper-bound is then simply:
\begin{eqnarray}
{\cal B}_1 \sim \frac{1}{c_1} \sqrt{\frac{\Pr \, \Ra}{{\cal C}}} \, .
\end{eqnarray}
Laboratory experiments are typically run in the range of absorption length $\ell \ll 1$: even though the dimensional absorption length $\ell_0$ is chosen to be greater than the boundary layer thickness, it remains much smaller than the height of the domain. In that limit, the constant $c_1$ is approximately given by $c_1=1/2+O(\ell)$. In the asymptotic limit $\Ra \to \infty$ for fixed $\ell \ll 1$, the upper-bound thus takes the compact asymptotic form:
\begin{eqnarray}
{\cal B}_1 \sim 2 \sqrt{\frac{\Pr \, \Ra}{{\cal C}}} \, . \label{asymptB}
\end{eqnarray}

\subsection{Restricting attention to fully turbulent solutions}

An interesting aspect of the upper bound (\ref{asymptB}) is that it allows to readily characterize the behavior of families of fully turbulent solutions, as defined at the outset. Of course, we have not proven that such families of solutions exist. However, one can assume that such a family of solutions exists and investigate the scaling behavior of the associated  Nusselt number. The fully turbulent family of solutions is characterized by a finite nonzero limit of the dissipation coefficient for asymptotically large forcing (asymptotically large Reynolds number, $\RaP$ and $\Ra$), see (\ref{eq:FTassumption}). In that limit, the upper-bound (\ref{asymptB}) behaves as $\sqrt{\Pr \, \Ra}$, i.e., it follows the so-called ultimate regime of thermal convection. Experimental studies report a scaling exponent compatible with this scaling behavior in $\Ra$, while an extensive three-dimensional numerical study has shown that the scaling behavior in both $\Ra$ and in $\Pr$ agrees well with the ultimate scaling when $\Pr \ll 1$, the behavior in $\Pr$ being different for $\Pr \gtrsim 1$. In other words, this upper bound appears to be sharp in Rayleigh number when compared to the available experimental and 3D numerical data.

However, we have also reported analytical and 2D numerical solutions of CISS that exceed the ultimate scaling behavior, with the Nusselt number increasing linearly in $\Ra$. The upper-bound (\ref{asymptB}) immediately indicates that these solutions cannot be fully turbulent, and indeed they appear to be extremely laminar, see~\cite{miquelPRF19}. In the following subsection, we derive an upper bound that applies to arbitrary flow solutions (including both laminar and turbulent ones).

\subsection{Upper bound for arbitrary flow solutions}

We now wish to derive an upper bound that holds for any flow solution, i.e., we stop restricting attention to fully turbulent branches of solutions. To wit, we derive an upper bound on the inverse dissipation coefficient ${\cal C}^{-1}$. The Poincar\'e inequality in the vertical direction yields:
\begin{eqnarray}
\la \bu^2 \ra c_3 \leq  \la (\partial_z \bu)^2 \ra \leq \la |\bnabla \bu|^2 \ra =  \frac{\cal C}{\Pr} \la \bu^2 \ra^{3/2} \, ,
\end{eqnarray}
where $c_3=\pi^2$ for no-slip top and bottom boundaries and $c_3=\pi^2/4$ for a free-slip top boundary.
The inequality above yields:
\begin{eqnarray}
\frac{\Pr}{\cal C} \leq \frac{\la \bu^2 \ra^{1/2}}{c_3} \label{tempboundC}
\end{eqnarray}
Using the intermediate steps in (\ref{tempw}), together with $\la w^2 \ra < \la \bu^2 \ra$, we obtain:
\begin{eqnarray}
\la \bu^2 \ra  \leq  \frac{\Pr \, \Ra}{\cal C} \, ,
\end{eqnarray}
which we substitute into (\ref{tempboundC}) to get:
\begin{eqnarray}
 \frac{\Pr }{\cal C} \leq \frac{\Ra}{c_3^2} \, .
\end{eqnarray}
Inserting this inequality in the expression of the bound ${\cal B}_1$ leads to the following upper bound ${\cal B}_2$:
\begin{eqnarray}
\Nu \leq {\cal B}_2 = \frac{1}{4} \left[ \frac{c_2}{c_1} + \sqrt{ \left(\frac{c_2}{c_1}  \right)^2 + \frac{4 \, \Ra}{c_1 c_3 }}  \right]^2 \, . \label{boundB1}
\end{eqnarray}
The upper bound ${\cal B}_2$ does not involve the dissipation coefficient anymore. We are interested in the large-$\Ra$ low-$\ell$  asymptotic behavior of the bound, given by:
\begin{equation}
{\cal B}_2 \sim \frac{2}{c_3} \, \Ra \, = \left\{ \begin{matrix}
 \frac{2}{\pi^2} \, \Ra \ & \text{for a no-slip top boundary     } \, , \\
 \frac{8}{\pi^2} \, \Ra \ & \text{for a free-slip top boundary} \, .
\end{matrix}   \right. \label{asymptB2}
\end{equation}
This upper bound is similar to the one derived in Ref.~\cite{miquelPRF19}\cor{, with the same scaling in $\Ra$, but a prefactor much better-suited to the source/sink function (\ref{adimS}): the prefactor of the upper bound in Ref.~\cite{miquelPRF19} diverges as $\ell \to 0$, whereas the prefactor of ${\cal B}_2$ reaches a finite limit as $\ell \to 0$, see (\ref{asymptB2}). The upper bound} scales linearly in $\Ra$, which is precisely the scaling behavior of an exact analytic laminar solution derived in Ref.~\cite{miquelPRF19} for free-slip top and bottom boundaries. \cor{The scaling behaviour of the upper bound is thus sharp for free-slip top and bottom boundaries. We believe it is probably also sharp for no-slip boundary conditions, but that remains to be proven. One way to prove this would be to adapt the analytic flow solution in Ref.~\cite{miquelPRF19} to no-slip boundaries instead of stress-free boundaries (see Refs.~\cite{Jimenez1987,Chini2009,Waleffe2015,Wen2020} for the computation of steady convective flows with various boundary conditions).}


To summarize this section, the Nusselt number of any flow solution cannot increase faster than $\Ra$, while the Nusselt number
associated with a turbulent family of solutions cannot increase faster than $\sqrt{\Ra}$. In other words, any scaling exponent $\gamma$ larger than $1/2$ is necessarily associated with non-turbulent solutions, according to the definition given at the outset. In the following, we characterize the dissipation coefficient in experimental and numerical realizations of CISS to assess the fully turbulent nature of the flow.

\section{Assessing the fully turbulent nature of the flow from experimental and DNS data \label{sec:DNSExp}}

\begin{figure}
    \centering
    \includegraphics[scale=0.3]{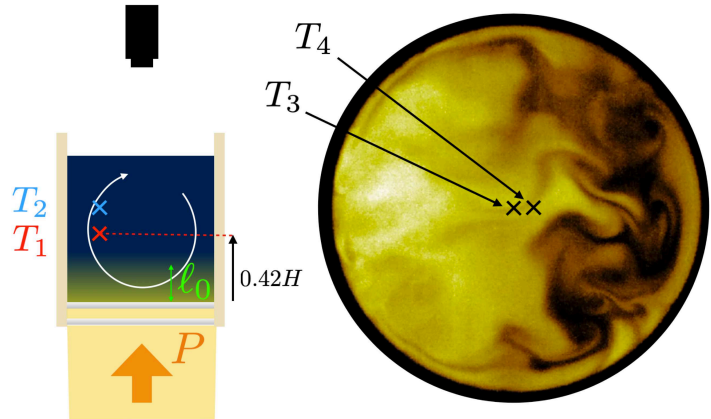}
    \caption{{\bf Left:} Schematic of the experimental setup with the thermal camera imaging the top free surface. $P$ denotes the total heat flux radiated from the spotlight in the form of visible light. $T_1$ and $T_2$ are the two vertically aligned temperature probes. {\bf Right:} snapshot of the temperature field on the top free-surface (lighter color for warmer temperature). From such snapshots we extract two local temperature signals at locations indicated by $T_3$ and $T_4$.}
    \label{fig:schema_exp}
\end{figure}

We now turn to numerical and experimental realizations of CISS. The goal is to extract the dissipation coefficient ${\cal C}$ and assess whether  the flow is fully turbulent \cor{according to (\ref{eq:FTassumption}).}

\subsection{Laboratory experiments\label{sec:Exp}}

\begin{figure}
    \centering
    \includegraphics[width=0.65\textwidth]{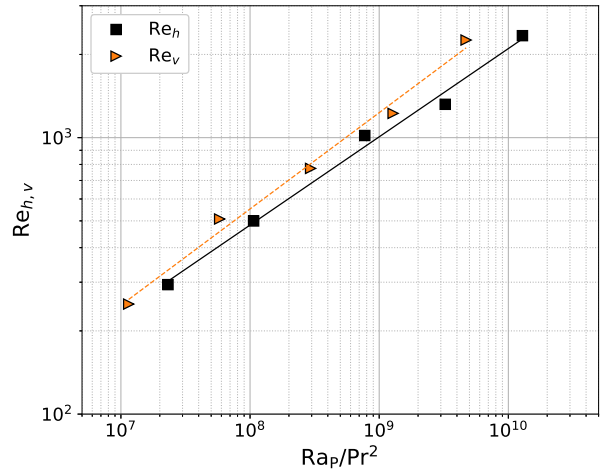}
    \caption{Experimental estimates of the Reynolds number as functions of $\mathrm{Ra_P Pr^{-2}}$ for $\ell=\ell_0/H=0.048$. The dashed line is $\mathrm{Re}=0.90\left(\mathrm{Ra_P\prandtl^{-2}}\right)^{0.35}$, while the straight line is $\mathrm{Re}=1.35\left(\mathrm{Ra_P\prandtl^{-2}}\right)^{0.32}$. \cor{$\mathrm{Re_h}$ (resp. $\mathrm{Re_v}$) refers to the Reynolds number inferred from the horizontal (resp. vertical) velocity estimate.}}
    \label{fig:ReRaP_exp}
\end{figure}

The experimental setup, sketched in Figure \ref{fig:schema_exp}, is described extensively in Ref.~\cite{lepotPNAS18}. We briefly recall its main characteristics. A cylindrical cell of radius $R=10$~cm is filled with a homogeneous mixture of water and carbon black dye, up to a height $H$ ranging from $4$ to $18$~cm. A powerful spotlight shines at the tank from below, the latter having a transparent bottom boundary. Absorption of light by the fluid leads to an internal source of heat, the magnitude of which decreases exponentially with height over a scale $\ell_0$ (Beer-Lambert's law). The absorption length $\ell_0$ can be tuned through the concentration of dye: for high dye concentration $\ell_0$ is much smaller than the thermal boundary layer thickness and we recover a standard RB-like boundary-heated configuration. By contrast, for low dye concentration $\ell_0$ is larger than the boundary layer thickness. The internal heating then bypasses the boundary layers and leads to the diffusivity-free mixing-length scaling regime of turbulent heat transport~\cite{lepotPNAS18,bouillautJFM19}. As one cannot cool down the fluid using a cold plate without inducing a heat-transport-restricting upper boundary layer, we resort to `secular heating' instead: in the absence of cooling and with insulating boundaries, the volume-averaged temperature grows linearly in time. However, the temperature difference between any two points inside the tank is governed by the equations of Boussinesq convection driven by the radiative heat source and effectively cooled at an equal and opposite rate by a uniform internal heat sink (see Ref.~\cite{lepotPNAS18} for details). This combination of radiative heating decreasing exponentially with height together with uniform effective cooling at an equal and opposite rate is the rationale behind the source/sink term in (\ref{eq:HE}). After a transient, a statistically steady vertical temperature drop arises.

The fluid being opaque, we cannot resort to standard optical velocimetry techniques to access the Reynolds number~\cite{Xin,Qiu2001,Xia2003,Xi,Sun,Liot}. We thus infer the local velocity from the temporal correlation between two neighboring temperature probes~\cite{chavannePoF01,castaingJFM89,Wu90,Wu92}. A vertical velocity estimate $U_v$ si inferred from the vertically aligned probes $T_1$ and $T_2$ \cor{in} Figure \ref{fig:schema_exp} (located at mid-radius, $0.42 H$ above the bottom boundary and $0.5$ cm apart, except for the highest $\RaP$ where they are $1$ cm apart). We compute the correlation function between the two probes, before dividing the spacing between the probes by the time-lag associated with the maximum of the cross-correlation function. This gives access to the characteristic velocity of a fluid parcel travelling from one probe to the other. In a similar fashion, we estimate the characteristic horizontal velocity $U_h$ through the cross-correlation between two virtual probes $T_3$ and $T_4$ (consisting of squares of four pixels each, $1$ cm apart) extracted from the thermal images, see Figure~\ref{fig:schema_exp}.



We build the inferred Reynolds number $\mathrm{Re_{h,v}}={U_{h,v}H}/{\nu}$ and plot them in Figure~\ref{fig:ReRaP_exp} as functions of $\RaP \prandtl^{-2}$. They are compatible with a power-law behavior ${\mathrm{Re_{h,v}}}\sim\left(\mathrm{Ra_P\prandtl^{-2}}\right)^{\beta}$, with $\beta=0.32$ and $\beta=0.35$ for $\mathrm{Re_h}$ and $\mathrm{Re_v}$, respectively. These values of $\beta$ should be compared to the laminar vs. turbulent values of this exponent: a fully turbulent flow that satisfies (\ref{eq:FTassumption}) is associated with $\beta=1/3$. By contrast, a laminar flow dissipating energy at the large scale $H$ would be characterized by a dissipation coefficient that is inversely proportional to the Reynolds number, hence $\beta=1/2$. The data in Figure~\ref{fig:ReRaP_exp} thus clearly discard the laminar scaling, while being compatible with the fully turbulent one within measurement accuracy. The \cor{measurement} accuracy is insufficient to characterise the scaling behavior of the dissipation coefficient more precisely, and we turn to numerical data instead.

\begin{figure}
    \centering
    \includegraphics[height=0.8\textwidth]{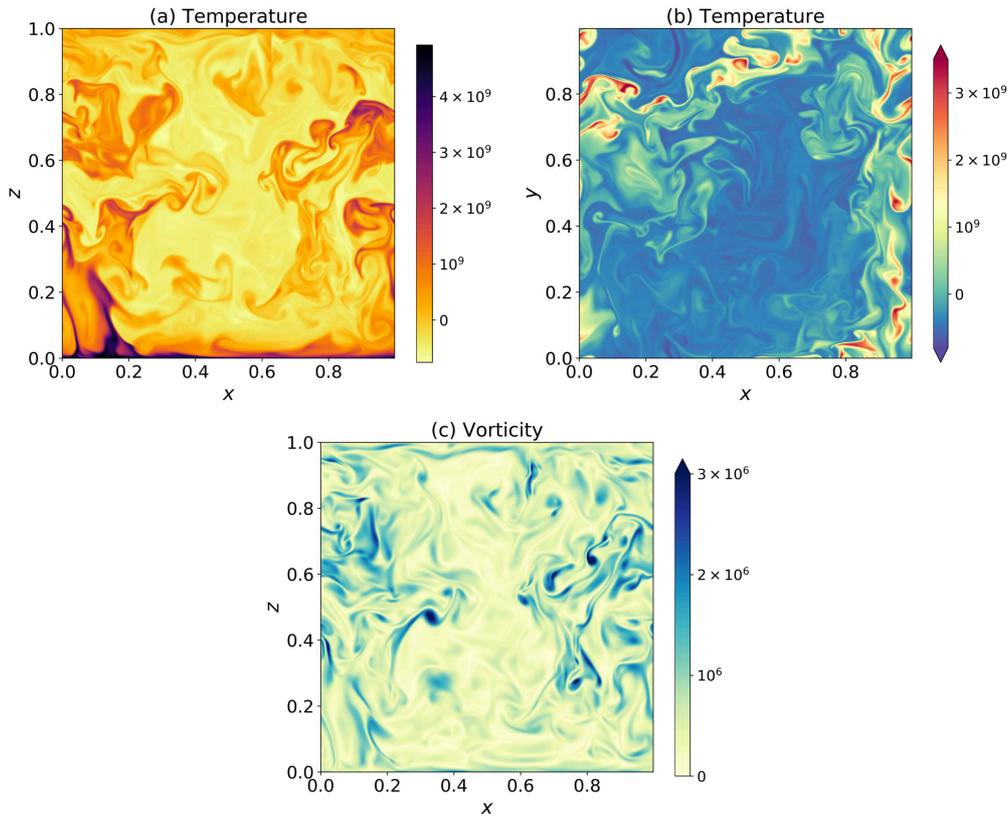}
    \caption{Snapshots of a turbulent flow computed numerically for $\RaP=10^{12}$\cor{, $\prandtl=7$ and $\ell=0.048$}. (a) Vertical slice of temperature. (b) Horizontal slice of temperature at $z=0.25$. (c) Vertical slice of vorticity $\sqrt{\left|\boldsymbol{\nabla}\times \bu \right|^2}$.}
    \label{fig:jolies_images}
\end{figure}

\subsection{Direct Numerical Simulations}

\begin{figure}
    \centering
    \includegraphics[width=0.6\textwidth]{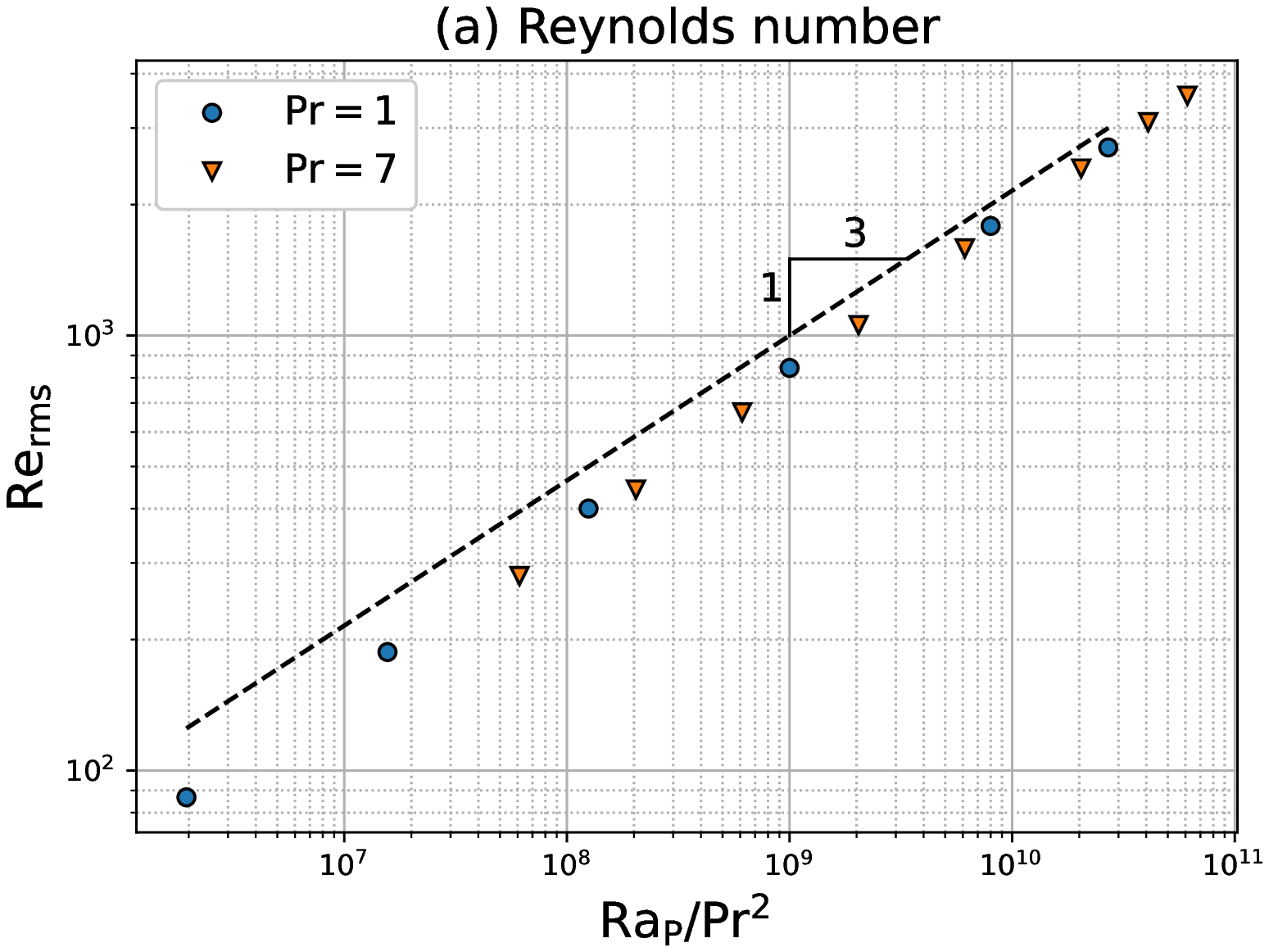}
    \includegraphics[width=0.6\textwidth]{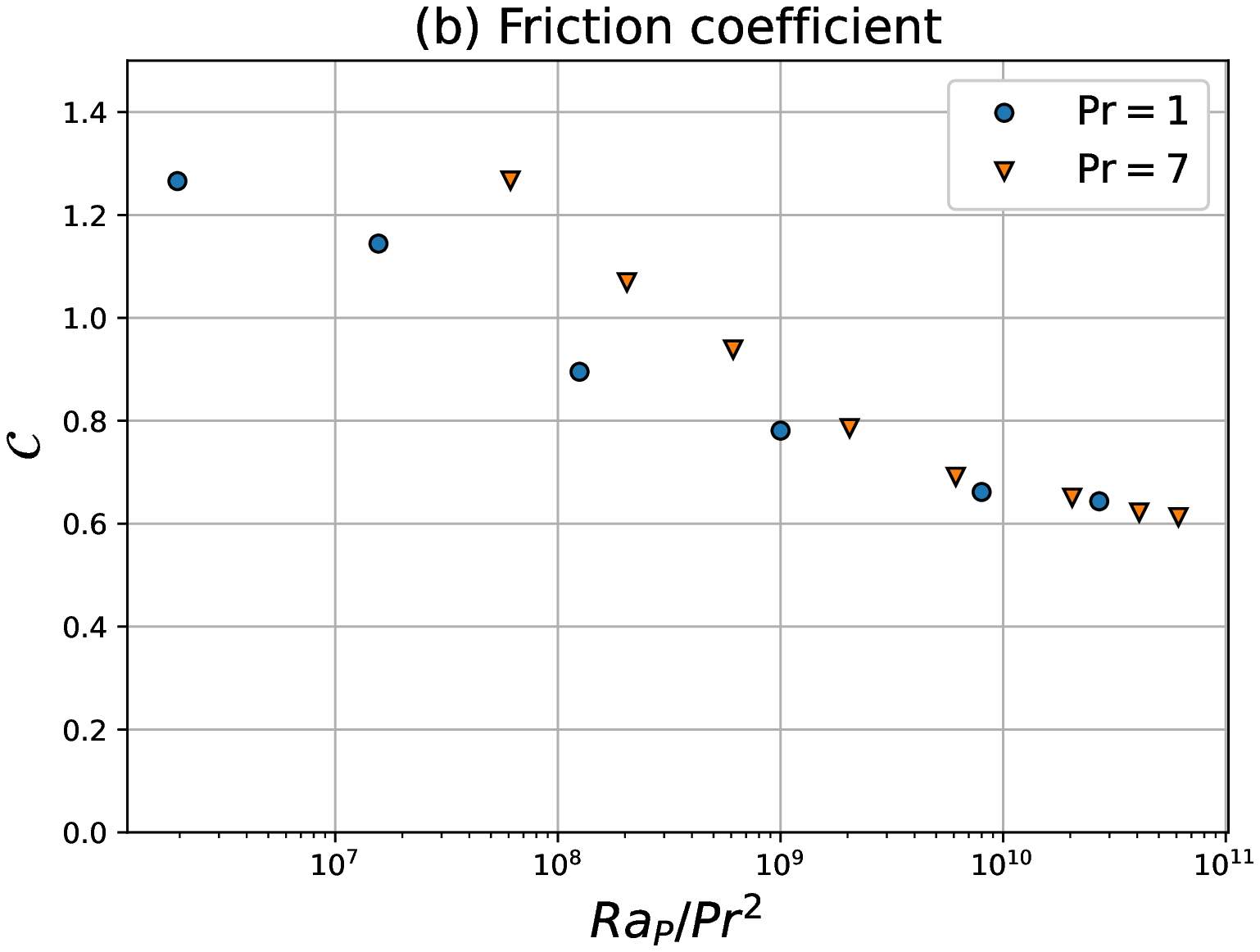}
    \caption{(a) Reynolds number, defined in equation~(\ref{def:ReRms}) and (b) Dissipation coefficient, defined in equation~(\ref{def:C}), obtained by DNS and plotted as functions of the control parameter $\RaP/\prandtl^2$. \cor{$\bullet$: $\prandtl=1$ and $\ell=0.024$; $\triangledown$: $\prandtl=7$ and $\ell=0.048$.}  \label{fig:C_DNS}}
\end{figure}

We compute numerical solutions to the governing equations (\ref{eq:dimless_governing}) using Coral, a pseudo-spectral, scalable, time-stepping solver for differential equations~\cite{miquelJOSS}. The computational domain is a unit cube $(x,y,z)\in [0,1)\times[0,1)\times [0,1]$ with periodic boundary conditions along the horizontal directions $(x,y)$. We impose impermeable and thermally insulating boundary conditions along surfaces $z=0$ and $z=1$, see equations (\ref{eq:impermeable_insulating}). We model the solid bottom and the free surface of the experiment by imposing no-slip boundary conditions at $z=0$ (\ref{eq:noSlip_bottom}) and free-slip boundary conditions at $z=1$ (\ref{eq:stressFree_top}). In Coral, these boundary conditions are imposed through basis recombination, i.e., by expanding the variables on bases of functions obtained as tensor products of Fourier modes along the horizontal and suitable linear combinations of Chebyshev polynomials, each of which obeys the boundary conditions along $z$ (see, e.g., Ref.~\cite{boydBOOK}). Simulations reported in the present manuscript use the second order semi-implicit time-stepping scheme of~\cite{ascherAMN97}. Finally, the divergence-free constraint is readily imposed by expressing the solenoidal velocity field $\bu$ in terms of velocity potentials $\phi$ and $\psi$, and horizontally-invariant mean flows $U$ and $V$:
\begin{equation}
\bu = \boldsymbol{\nabla} \times \psi(x,y,z,t)\, \ez + \boldsymbol{\nabla} \times  \boldsymbol{\nabla} \times \phi (x,y,z,t)\, \ez + U(z,t)\,\ex + V(z,t)\,\ey \, .
\end{equation}

\cor{The data consist in sweeps of the flux-based Rayleigh number $\RaP$ for two values of the Prandtl number: the canonical case $\prandtl=1$ with $\ell=0.024$, and the value $\prandtl=7$ of laboratory experiments using water as the working fluid, together with a dimensionless absoprtion length $\ell=0.048$}. 
We illustrate in Figure~\ref{fig:jolies_images} the turbulent flow obtained \cor{for $\RaP=10^{12}$, $\Pr=7$ and $\ell=0.048$}.  

For each set of parameters, the flow is computed from an initial condition taken as either random small-amplitude noise, or the final state of a run with neighboring control parameters. We let the flow equilibrate and reach a statistically stationary regime during which some velocity and temperature diagnostics are computed. In the following, we focus specifically on the dissipation coefficient ${\cal C}$, the Rayleigh and Nusselt numbers -- using the definitions (\ref{eq:defRamax}-\ref{eq:defNumax}) for ease of comparison with experiments -- and the Reynolds number $\ReRms$ based on the height of the domain and the root-mean-square velocity. In terms of the dimensionless variables, $\ReRms$ is defined as:
\begin{equation}
    \ReRms = \frac{\sqrt{\la \left|\bu\right| ^2 \ra}}{\prandtl} \label{def:ReRms}\,. 
\end{equation}

%

We plot \cor{in} figure~\ref{fig:C_DNS} both the Reynolds number $\ReRms$ and the dissipation coefficient ${\cal C}$, as functions of $\RaP/\prandtl^2$. Both the $\prandtl=1$ and $\prandtl=7$ datasets behave similarly and indicate a tendency towards turbulent dissipation for our most turbulent flows, corresponding to $\ReRms\gtrsim2000$. Indeed, these data points seem to asymptote the scaling law $\ReRms\sim \left(\RaP / \prandtl^2\right)^{1/3}$, and the dissipation coefficient $\mathcal{C}$ for both $\prandtl=1$ and $\prandtl=7$ seems to saturate to a limiting value ${\cal C}_\infty \simeq0.6$. Data at even higher Reynolds number \cor{may} be necessary to reach a definitive conclusion. However, the present data strongly suggest that CISS enters \cor{a fully turbulent regime characterized by anomalous dissipation for $\ReRms\gtrsim2000$, in line with (\ref{eq:FTassumption}), which} justifies the applicability of both bounds ${\cal B}_1$ and ${\cal B}_2$ in section~\ref{sec:VelInf} to this flow.

\section{Discussion \label{sec:Disc}}

\begin{figure}
    \centering
    \includegraphics[width=\textwidth]{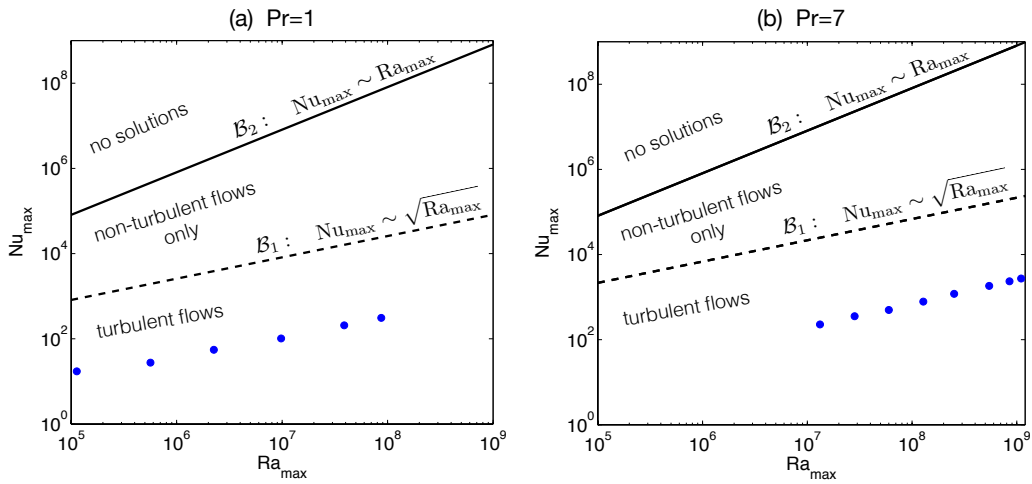}
    \caption{Numerical data (blue symbols) together with upper bounds ${\cal B}_1$ and ${\cal B}_2$ in the plane ($\mathrm{Ra}_\text{max},\mathrm{Nu}_\text{max}$) for (a) $\text{Pr}=1$ and (b) $\text{Pr}=7$. The upper bound ${\cal B}_1$ over turbulent branches of solutions is represented using the numerically determined asymptotic value of the dissipation coefficient ${\cal C}_\infty \simeq 0.6$. \cor{The absorption length used in the DNS is $\ell=0.024$ for panel (a) and $\ell=0.048$ for panel (b).}\label{fig:bilan}}
\end{figure}

Focusing on convection induced by heat sources and sinks, we have derived \cor{lower bounds on the emergent root-mean-square temperature in terms of the flux-based Rayleigh number. To make better contact with the vast literature on thermal convection, we have expressed these bounds in terms of the emergent temperature-based Rayleigh and Nusselt numbers. As argued at the outset, the definitions (\ref{eq:defRamax}-\ref{eq:defNumax}) are probably best-suited for comparison with experimental results, and we provide in Figure~\ref{fig:bilan} a summary of the bounds in the plane $(\Ramax,\Numax)$.} The upper bound (\ref{asymptB}) on the Nusselt number scales as the square-root of the Rayleigh number. It applies only to turbulent branches of solutions, according to the definition (\ref{eq:FTassumption}) given at the outset and assuming that such branches of solutions exist. By contrast, the best upper bound (\ref{asymptB2}) over all flow solutions scales linearly in Rayleigh number, a behavior associated with laminar analytic flow solutions such as the one derived in Ref.~\cite{miquelPRF19}.

We then turned to numerical and experimental data to validate the existence of turbulent branches of solutions. The numerical data indeed point to a finite limiting value of the dissipation coefficient as the flux-based Rayleigh number -- and thus the Reynolds number -- increases. We stress the fact that the validation of the fully turbulent assumption (\ref{eq:FTassumption}), when combined with the `ultimate' scaling-law for heat transport $\Nu \sim \sqrt{\Ra \prandtl}$, leads to Spiegel's free-fall scaling-law for the velocity field $\mathrm{Re} \sim \sqrt{\Ra/\prandtl}$, where the Reynolds number $\mathrm{Re}$ is based on the height of the domain and the root-mean-square velocity. In other words, using CISS we can validate both Spiegel's prediction of a diffusivity-free regime for the heat transport, $\Nu\sim \sqrt{\Ra \prandtl}$, and the free-fall scaling assumption that underpins it~\cite{Spiegel71}.

The upper bounds derived in the present study provide a useful point of view to discuss the proposed scaling theories for CISS. For instance, in Lepot et al.~\cite{lepotPNAS18} we report a scaling exponent $\gamma \simeq 0.55$ for the scaling-law $\Nu \sim \Ra^\gamma$ at $\prandtl = 7$. Assuming that the flow becomes fully turbulent at large driving amplitude, the slight departure of the measured exponent from $1/2$ must be attributed to finite-Reynolds-number effects. As a matter of fact, we showed in Lepot et al. using DNS that the scaling exponent $\gamma$ is indeed much closer to $1/2$ for $\prandtl=1$ than for $\prandtl=7$, the former corresponding to reduced viscous effects. The upper bounds (\ref{asymptB}) and (\ref{asymptB2}) are also fully compatible with the scaling predictions that we put forward in Refs.~\cite{bouillautJFM19} and~\cite{miquelJFM20} with regard to the dependence of the Nusselt number on $\Ra$, $\prandtl$ and the dimensionless absorption length $\ell$. By contrast, the upper bounds challenge some scaling predictions recently put forward in Ref.~\cite{creyssels2020} for CISS: out of the five scaling regimes proposed in Ref.~\cite{creyssels2020}, three correspond to a scaling exponent $\gamma>1$ and are thus discarded by the general upper bound (\ref{asymptB2}). The remaining two scaling regimes have a scaling exponent $\gamma>1/2$: they appear to violate the upper-bound (\ref{asymptB}) on fully turbulent branches of solutions and thus require a dissipation coefficient that vanishes asymptotically at large driving amplitude. It thus appears that care must be taken when applying the intuition gathered from the Rayleigh-B\'enard setup to CISS. The present data indicate that, to some extent, CISS behaves in a much simpler fashion than RB convection: as in most instances of fully turbulent flows, the speed, the dissipated power and the transport properties seem to become independent of molecular diffusivities for large-enough driving amplitude. In other words, in line with Spiegel's intuition~\cite{Spiegel71} the convective flow achieves fully turbulent dissipation, the free-fall velocity scaling-law and the ultimate heat transport scaling-law.

\funding{This research is supported by the European Research Council under grant agreement FLAVE 757239. The numerical study was performed using HPC resources from GENCI-CINES and TGCC (grant 2020-A0082A10803 and grant 2021-A0102A10803).}

\bibliographystyle{rsta}
\bibliography{Proceedings_convection_NoDOI}

\end{document}